\documentclass[conference,10pt]{IEEEtran}
\ifCLASSINFOpdf
\else
\fi

\usepackage{notation}
\usepackage{epsfig}
\usepackage{color}
\usepackage{amsmath}
\usepackage{amssymb}
\usepackage{enumerate}
\usepackage{multirow}
\usepackage{bbm}

\title{Scaling Behavior of Convolutional LDPC Ensembles over the BEC}

\begin{document}
%\title{Scaling Behavior of Convolutional LDPC Ensembles over the BEC}

%\title{\Large{Scaling Behavior of Convolutional LDPC Ensembles over the BEC}}
%\author{
%\IEEEauthorblockN{
%Pablo M. Olmos\IEEEauthorrefmark{1} and
%R\"udiger Urbanke\IEEEauthorrefmark{2}
%}
%
%\IEEEauthorblockA{\IEEEauthorrefmark{1}
%Teoria de la senal y communicaciones, \\ Universidad de Sevilla 
%\\ Sevilla, Spain \\ Email: olmos@us.es}
%
%\IEEEauthorblockA{\IEEEauthorrefmark{2}School of Computer and Communication Sciences\\
%EPFL, Lausanne, Switzerland
%\\Email: ruediger.urbanke@epfl.ch}
%}
\author{
\authorblockN{Pablo M. Olmos}
\authorblockA{Departamento de Teor\'ia de la Se\~nal y Comunicaciones.\\
Universidad de Sevilla.\\
email:\tt olmos@us.es}
\and
%\authorblockN{Juan Jos\' e Murillo-Fuentes}
%\authorblockA{Departamento de Teor\'ia de la Se\~nal \\ y Comunicaciones\\
%Universidad de Sevilla.\\
%Email: murillo@us.es}
%\and
\authorblockN{R\"udiger Urbanke,
}
\authorblockA{School of Computer and Communication Sciences\\
EPFL, Lausanne, Switzerland
\\email:\tt ruediger.urbanke@epfl.ch}
}
%\author{Pablo M. Olmos and~R\"{u}diger Urbanke% <-this % stops a space
%%\thanks{I would like to thank Professor R\"{u}diger Urbanke and all the members of the Information Processing Group at EPFL for the opportunity I have given to work and learn from them.}% <-this % stops a space
%\thanks{Pablo M. Olmos is with the University of Sevilla, Spain. R\"{u}diger Urbanke is with the \'Ecole Polytechnique F\'ed\'erale de Lausanne, Switzerland. }% <-this % stops a space
%}

%\markboth{Information Processing Group Internal Report}%
%{Shell \MakeLowercase{\textit{et al.}}: Bare Demo of IEEEtran.cls for Journals}



\maketitle

\begin{abstract}
We study the scaling behavior of coupled sparse graph codes over the
binary erasure channel.  In particular, let $2L+1$ be the length of the
coupled chain, let $M$ be the number of variables in each of the $2L+1$
local copies, let $\ell$ be the number of iterations, let $P_{\text{b}}$
denote the bit error probability, and let $\epsilon$ denote the channel
parameter. We are interested in how these quantities scale when we let the
blocklength $(2L+1) M$ tend to infinity.  Based on empirical evidence
we show that the threshold saturation phenomenon is rather stable with
respect to the scaling of the various parameters and we formulate some
general rules of thumb which can serve as a guide for the design of
coding systems based on coupled graphs.  \end{abstract}


%\begin{IEEEkeywords}
%Convolutional LDPC codes, Spatial Coupling, Scaling Behavior.
%\end{IEEEkeywords}

% \fi
\IEEEpeerreviewmaketitle

\section{Introduction}
\let\thefootnote\relax\footnotetext{This work was supported by grant No. 200021-125347 of the Swiss
National Foundation and by Spanish government MEC TEC2009-14504-C02-\{01,02\} and Consolider-Ingenio 2010 CSD2008-00010).} 
\IEEEPARstart{S}{patially coupled} codes \cite{Kudekar10} provide an
entirely new way of approaching capacity. The basic phenomena can be
phrased as follows: an ensemble constructed by coupling a chain of
$(2L+1)$ regular $(l,r)$ low-density parity-check (LDPC) ensembles,
together with appropriate boundary conditions of the chain, exhibits
a belief propagation (BP) threshold close to the maximum-a-posteriori
(MAP) threshold of the regular $(l,r)$ ensemble.  This phenomenon is
known as \emph{threshold saturation} and it has been proved rigorously
for the binary erasure channel (BEC) in \cite{Kudekar10}.
It has also been observed empirically for a variety of other channels and other
graphical models in \cite{Tanner04,Lentmaier10,Kudekar10-2}.
\emph{Low-density parity-check convolutional} (LDPCC) ensembles, first
introduced in \cite{FelstromZ99}, are the best known example of spatially
coupled codes.  In \cite{Lentmaier09}, the authors reformulate
LDPCC ensembles in terms of protographs.  The BP threshold for these codes
is computed using density evolution (DE) in \cite{Lentmaier10, Kudekar10}
and it is conjectured that they achieve capacity universally across the
set of binary-input memoryless output-symmetric channels \cite{Kudekar10}.

It is probably fair to state that by now the asymptotic
performance of spatially coupled LDPC codes is well understood.  However,
much less is known about their scaling behavior \cite{Kudekar10}.
For instance, the DE analysis of LDPCC codes typically assumes that $L$ is
kept fixed while $M$ tends to infinity. But, does the threshold saturation
phenomena happen even if $L$ grows as a function of $M$? In this work,
we analyze the finite-length performance of LDPCC codes and we study
how it scales with the coupling dimensions $M$ and $L$. Our empirical
observations indicate that the threshold saturation phenomenon happens
even when $L$ grows considerably faster than $M$, which indicates that
the threshold saturation phenomenon is very robust.  
From our simulation results we synthesize some general design rules for these codes.
In particular, if the code-length is bounded, how should we chose $L$
and $M$ to have the best performance? And how does this choice affect
the decoder complexity (in terms of average number of iterations)? These
questions, among others, are of significantly practical importance.

The study of the finite-length behavior of LDPCC codes is
augmented by analyzing their error floor \cite{Urbanke08-2}.
In \cite{Kudekar10,Lentmaier10-2}, the authors prove that the
minimum distance of LDPCC codes is a fraction of $M$. These studies
concern ``large'' weight codewords. We investigate the occurrence of
constant-sized codewords/stopping sets, and in particular their scaling.
We prove that the fraction of codes with no error floor is roughly
equal to $\exp(-c L/M^{l-2})$, where $c$ only depends on the rate of the
code. Hence, for sufficiently small ratios $L/M^{l-2}$, it is very easy
to expurgate the ensemble and to find codes with linear minimum distance.

%The rest of the paper is organized as follows. Section\SEC{Bpar}
%is devoted to introduce of basic notions of LDPCC codes. In
%Section\SEC{Iter}, we study the decoding complexity, focusing on the
%required number of iterations for the BP decoder and how it depends on
%the code parameters. Finally, in Sections \SEC{Thbeha}-\SEC{errorfloor}, the performance
%of LDPCC codes as function of $(M,L)$ is analyzed.

\section{Convolutional-like LDPC ensembles. Basic design parameters}\LABSEC{Bpar}
\color{black}
We define the LDPCC ensembles using protographs
\cite{Lentmaier09}.  We start from a collection of $(2L+1)$
regular $(l,r=kl)$ LDPC protographs with $k\in\mathbb{N}$ \cite{Thorpe03}
and so that $l$ is odd, as shown in Fig.~\FIG{Fig1} for $(l,r)=(3,6)$
and $L=9$. The regular $(l,r=kl)$ code is referred to
as the {\em underlying} code. The associated protograph has $k$ variable
nodes of degree $l$ so that, if $M$ is the total number of variables
per protograph, each variable node of the protograph represents $M/k$
variables in total. For instance, in Fig.~\FIG{Fig1}, each variable
node of the protograph represents $M/2$ variables. In the following,
we say that the LDPCC graph has $(2L+1)$ sections, one per protograph
in Fig.~\FIG{Fig1}.

Let us now define the \emph{coupled protograph}.  This graph is
constructed by \emph{spatially coupling} the 
protographs in Fig.~\FIG{Fig1}: each variable node is
connected to its $\hat{l}$ check node neighbors  on the left and to its
$\hat{l}$ check node neighbors  on the right, where $\hat{l}=(l-1)/2$ \cite{Kudekar10}.
The coupled protograph is terminated by adding $\hat{l}$ extra check
nodes on each side. This process is illustrated in Fig.~\FIG{Fig2}
for the case $(l,r,L)=(3,6,9)$. In the termination procedure described,
the check nodes of lower degree on both sides provide better protection
for the connected variables.  However, there is a price to be payed for
this extra protection -- the rate is reduced with respect to the rate
of the underlying code.  The ensemble has  $n=M(2L+1)$ variable nodes
and $(2(L+\hat{l})+1)M/k$ check nodes and the design rate is:
\begin{equation}\LABEQ{ratered}
R(l,r=kl,L)=\frac{k-1}{k}-\frac{2\hat{l}}{k(2L+1)},
\end{equation}
where the first term is the rate of the underlying code
\cite{Kudekar10}. 
To generate a sample of the LDPCC ensemble we now ``lift'' the coupled
protograph in the same manner as this is done for regular ensembles
\cite{Thorpe03}: we make $(M/k)$ copies of the coupled
protograph and we connect them by picking for each ``edge bundle'' a
random permutation. In the following
we refer to the ensemble as the $(l,r,L,M)$ (convolutional) ensemble.

% Since the LDPCC ensemble is completely determined
%by the underlying code and the parameters $M$ and $L$, 

\begin{figure}
\centering
\includegraphics[width=5.75cm]{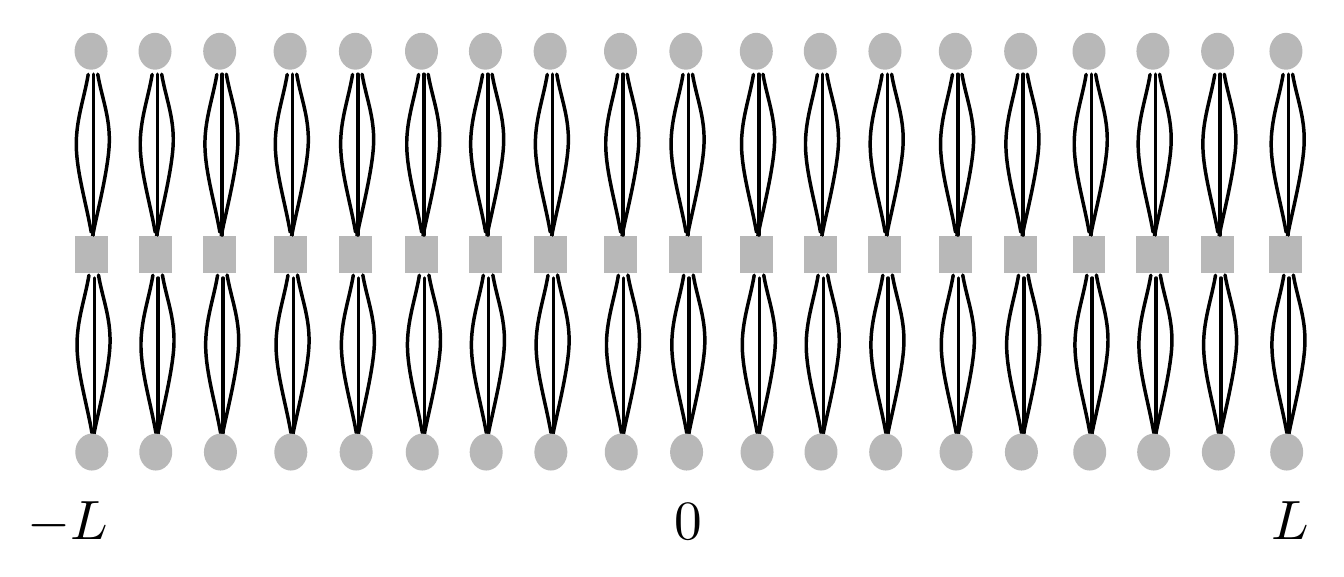} \caption{A
chain of $(2L+1)$ regular (3,6) non-interacting protographs for
$L=9$.}\LABFIG{Fig1} \end{figure}

\begin{figure}
\centering
\includegraphics[width=5.75cm]{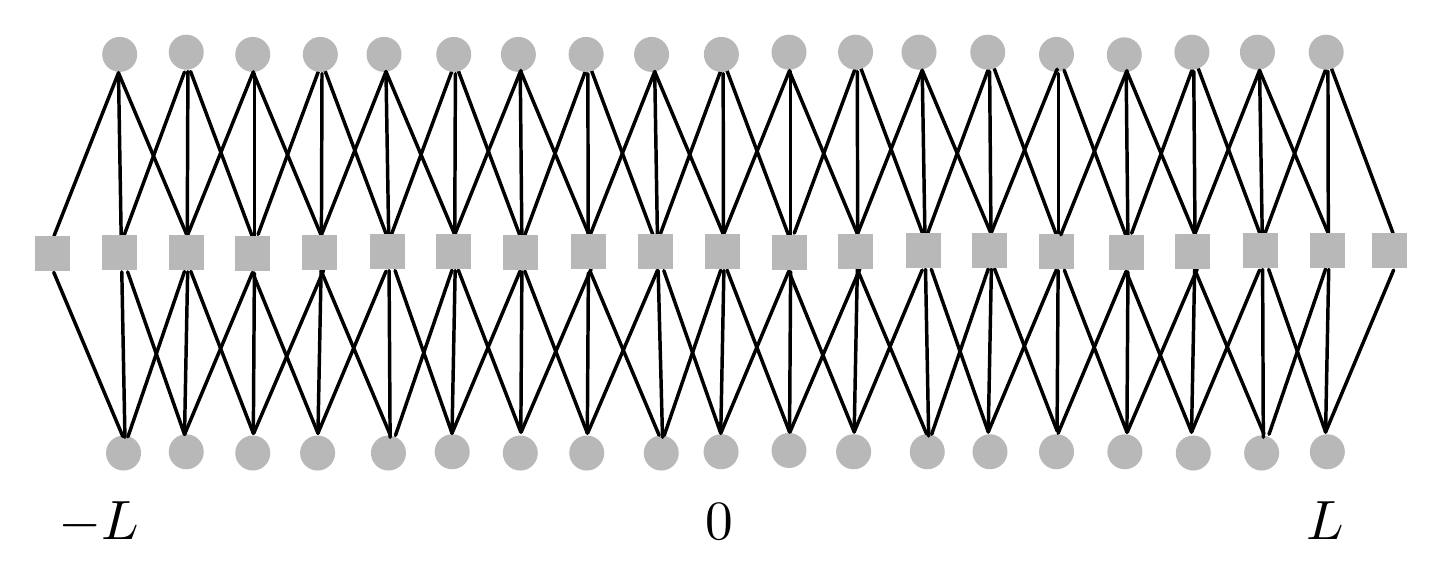} \caption{Coupled
protograph created from a chain of $(2L+1)$ regular (3,6) protographs
for $L=9$.}\LABFIG{Fig2} \end{figure}

%Remember that, as
%discussed, the number of sections is $(2L+1)$ and that the number
%of variables per section is $2M$.
\subsection{Asymptotic analysis of the LDPCC ensemble}
The performance of spatially-coupled ensembles under BP decoding as $M$ goes to infinity is analyzed
in \cite{Lentmaier10,Kudekar10} using density evolution (DE)
\cite{Urbanke08-2}. This allows to compute the \emph{BP threshold},
which defines the limit of the decodable region.  Let us denote the
threshold for the BEC by $\epsilon^{\text{BP}}(l,r,L)$. We have
\begin{align}\LABEQ{l1}
&\lim_{\ell\rightarrow\infty}\lim_{M\rightarrow\infty} P_{b}^{\ell}(\epsilon,l,r,L,M)=0,
\;\epsilon<\epsilon^{\text{BP}}(l,r,L),
\end{align}
where $P_{b}^{\ell}(\epsilon,l,r,L,M)$ is the ensemble average bit error probability after $\ell$ decoding rounds:
\begin{equation}\LABEQ{Average}
P_{b}^{\ell}(\epsilon,l,r,L,M)=
\E_{\mathcal{C}\in(l,r,L,M)}[P_{b}^{\ell}(\epsilon,\mathcal{C})].
\end{equation}
Similarly, $P_{B}^{\ell}(\epsilon,l,r,L,M)$ denotes the ensemble average
block error probability.  One of the key results of the asymptotic
analysis of coupled codes is that $\epsilon^{\text{BP}}(l,r,L)$ is
``very close'' to $\epsilon^{\text{MAP}}(l,r)$, the MAP threshold of
the underlying regular ensemble \cite{Lentmaier10}.

\subsection{Finite-length scaling LDPCC codes} 
Finite-length scaling investigates the relationship between the
performance, the code parameters, and the decoding complexity.
%We define the pair $\Delta=\{n,\ell_{\max}\}$ as a basic complexity
%measure, where $n$ is the code-length and $\ell_{\max}$ is the maximum allowed
%number of iterations of the BP decoder. 
Any practical design of a LDPCC code starts from a set of constraints
on the code rate in \EQ{ratered}, the code length, and the number of decoding
iterations with the goal of finding the best choice of parameters.
To first order, one might wonder for which
scaling of $L$ with respect to $M$ the threshold saturation phenomenon
occurs. More precisely, if $L=f(M)$, for what functions $f(\cdot)$ does the limit
\begin{equation}\LABEQ{limitfM2}
\lim_{\ell\rightarrow\infty}\left(\lim_{M\rightarrow\infty} P_{B}^{\ell}\left(\epsilon,l,r,L=f(M),M\right)\right)
\end{equation}
converges to $0$ for all $\epsilon<\epsilon^{\text{BP}}(l,r,L) $ as stated
in \EQ{l1}?  In Section\SEC{Thbeha}, we investigate this question 
by testing the code performance for several scaling functions $f(M)$
and increasing $M$ values.

\section{Decoding complexity}\LABSEC{Iter}
A practical implementation of a message-passing decoder has to set the
number of iterations, call it $\ell_{\min}$, which ensures a reliable
decoding in most cases. To be precise, assume that we have to design
$\ell_{\min}$ so that the decoder succeeds with probability higher than
$\delta$.  In \cite{Lentmaier10}, this task is addressed via DE. 
We have empirically computed \emph{the
ensemble average distribution of the required number of iterations}. It
is defined as follows \cite {Urbanke08-2}(Chapter 3, Section 22):
\begin{align}\LABEQ{Dist}
&\varphi(\ell,\epsilon,L,M)=P_{B}^{\ell-1}(\epsilon,l,r,L,M)-P_{B}^{\ell}(\epsilon,l,r,L,M),
\end{align} 
for $\ell\geq1$.
%, where $P_{B}^{\ell}(\epsilon,l,r,L,M)$ is the ensemble
%average block error probability. 
Note that the associated cumulative function
\begin{align}\LABEQ{}
\Phi(\ell_{0},\epsilon,L,M)&=\sum_{\ell=1}^{\ell_{0}}\varphi(\ell,\epsilon,L,M)\nonumber\\
&=P_{B}^{0}(\epsilon,l,r,L,M)-P_{B}^{\ell_{0}}(\epsilon,l,r,L,M)\nonumber\\
%&=1-(1-\pe)^{n}-P_{B}^{\ell_{0}}(\epsilon,l,r,L,M)\nonumber\\
&\approx1-P_{B}^{\ell_{0}}(\epsilon,l,r,L,M),
\end{align}
provides the probability of successful decoding after $\ell_{0}$
iterations. Therefore, $\ell_{\min}$ is chosen so that
$\Phi(\ell_{\min},\epsilon,L,M)\geq\delta$.

It is clear that for $\epsilon < \epsilon^{\text{BP}}(l, r)$,
$\ell_{\min}$ is essentially independent of $L$ and has the same
distribution as the distribution for the regular $(l,r)$ code of length $M$.
%More precisely, in this regime, the distribution of
%$\varphi(\ell,\epsilon,L,M)$ converges with $L$ to that of the regular
%$(l,r)$ code.This is true since i
In this regime all sections can be
decoded at the same time and the effect of the boundary condition
vanishes once $L$ has become sufficiently large. It is easy to give a
coarse upper bound on how large $L$ has to be for this to be true.
%
%How large does $L$ have to be for this convergence
%to have essentially happened?  It is easy to give a coarse upper bound.
Fix the ``gap'' $\epsilon^{\text{BP}}(l, r)-\epsilon>0$. We can determine
via DE the required number of iterations for the $(l,r)$ ensemble to
bring down the error probability to a desired small value. Assume that
$L$ is large compared to this number of iterations.  Then the effect
of the boundary has not reached the middle section of the code by the
time it has essentially decoded. 
%We therefore see that in this regime
%the required number of iterations behaves essentially like the required
%number of iterations for the underlying code.  
%%Of course, this required
%%length increases as the gap to $\epsilon^{\text{BP}}(l, r)$ vanishes.

\begin{figure}[h]
\centering
%\begin{tabular}{cc}
%%\includegraphics[width=3.75cm]{Figures/L_20_M_512_Iter_5} & \includegraphics[width=3.75cm]{Figures/L_20_M_512_Iter_70}\\
%%(a) & (b)
%%\end{tabular}
\begin{tabular}{cc}
\includegraphics[width=3.5cm]{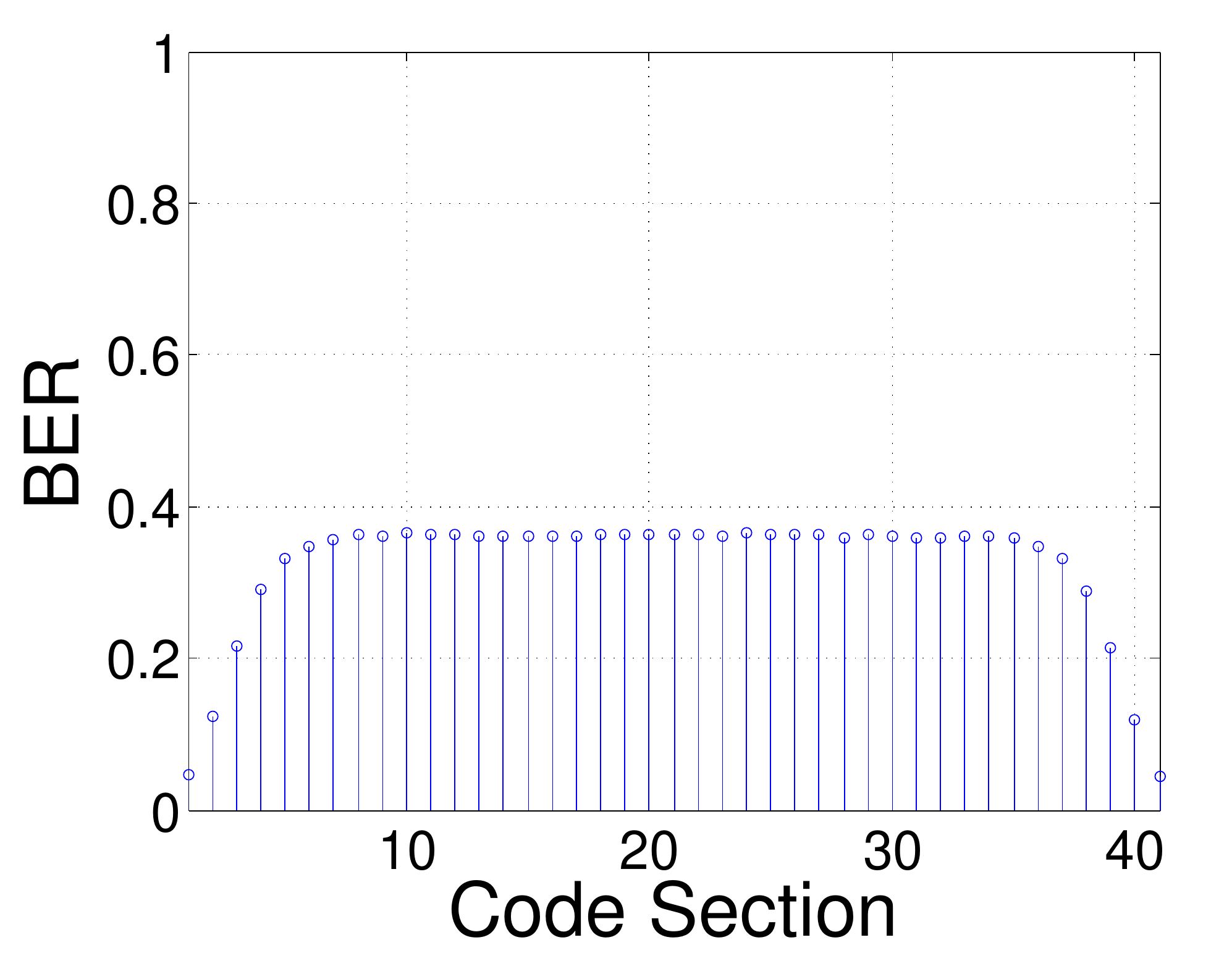} & \includegraphics[width=3.5cm]{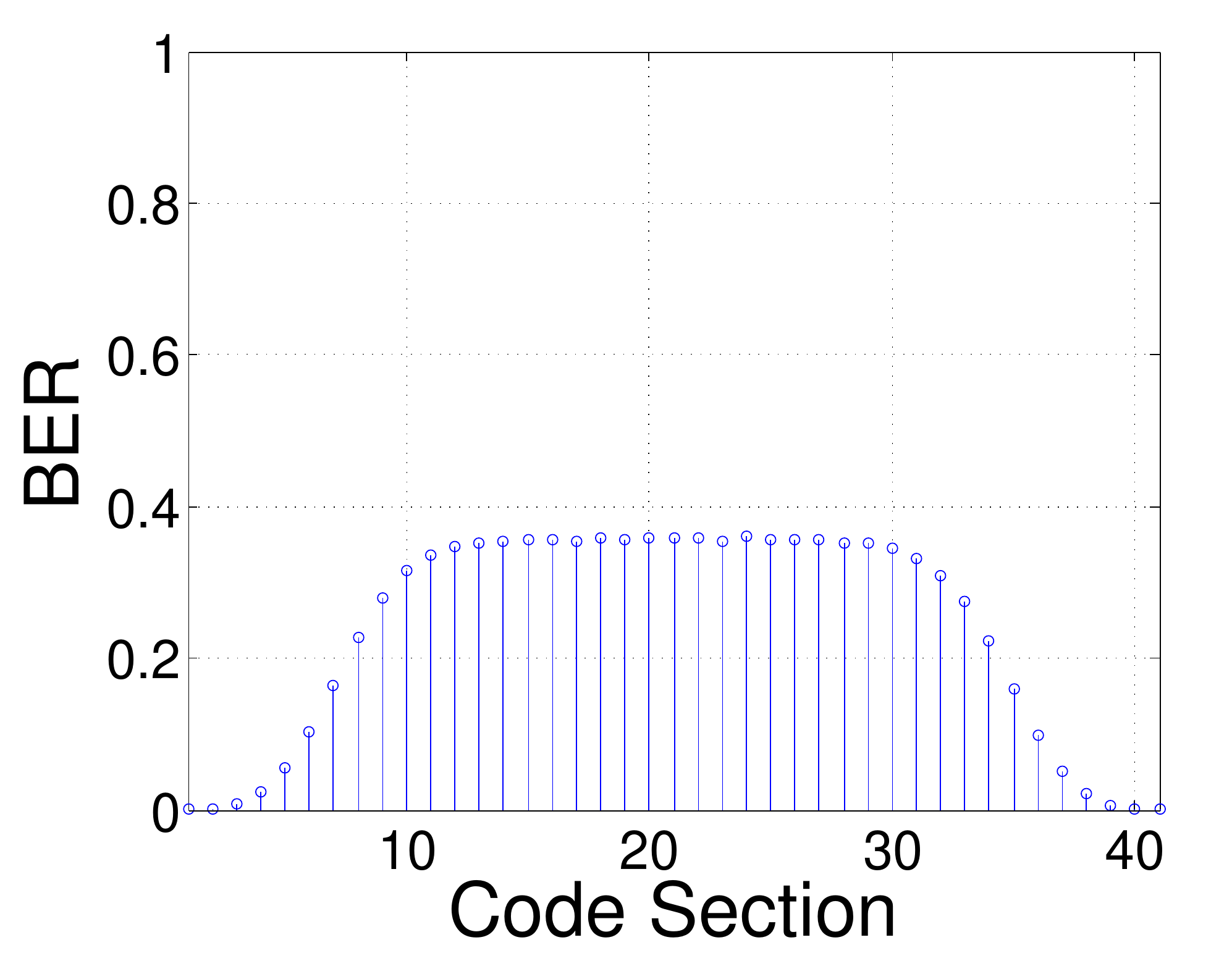}\\
(a) & (b)\\
\includegraphics[width=3.5cm]{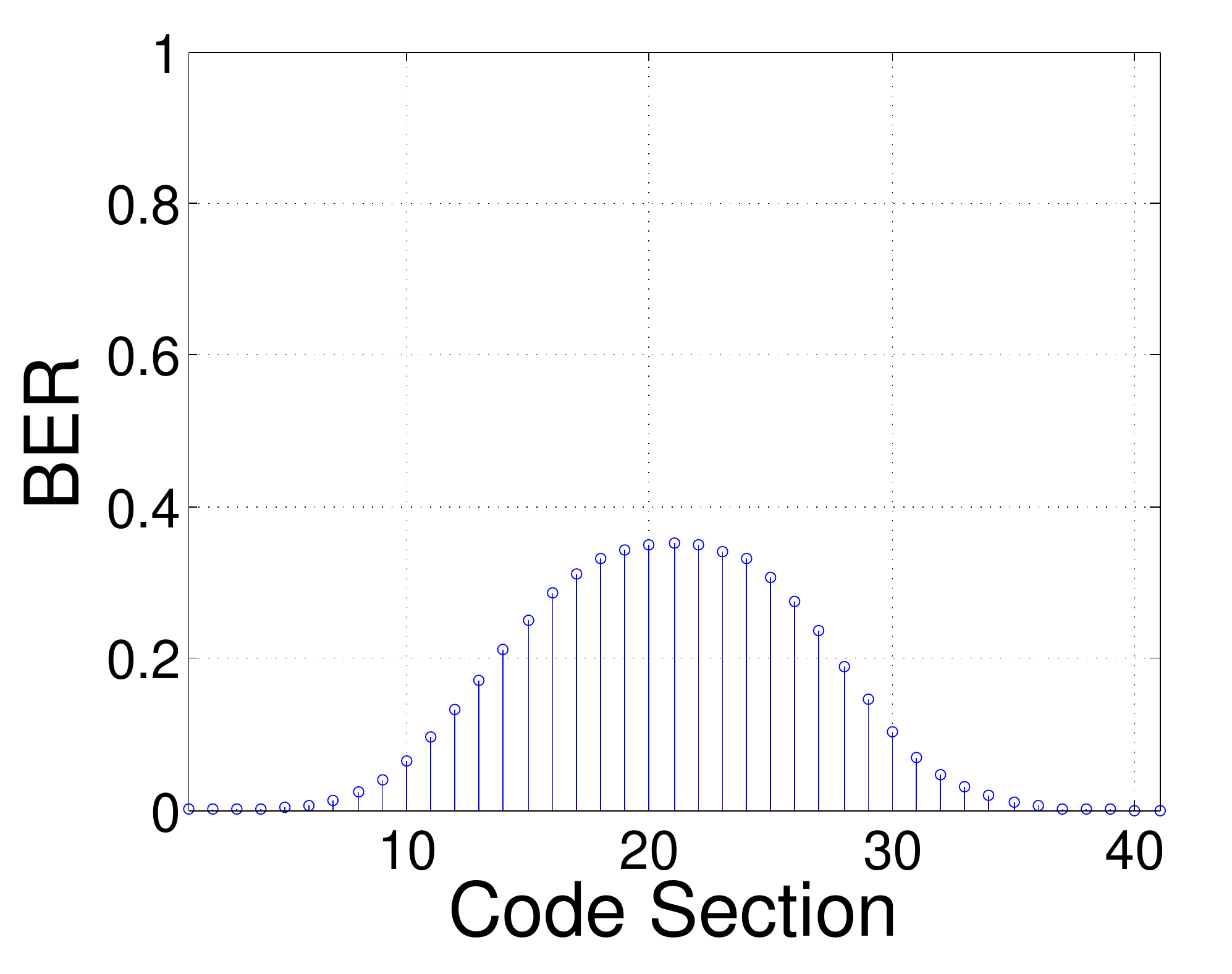} & \includegraphics[width=3.5cm]{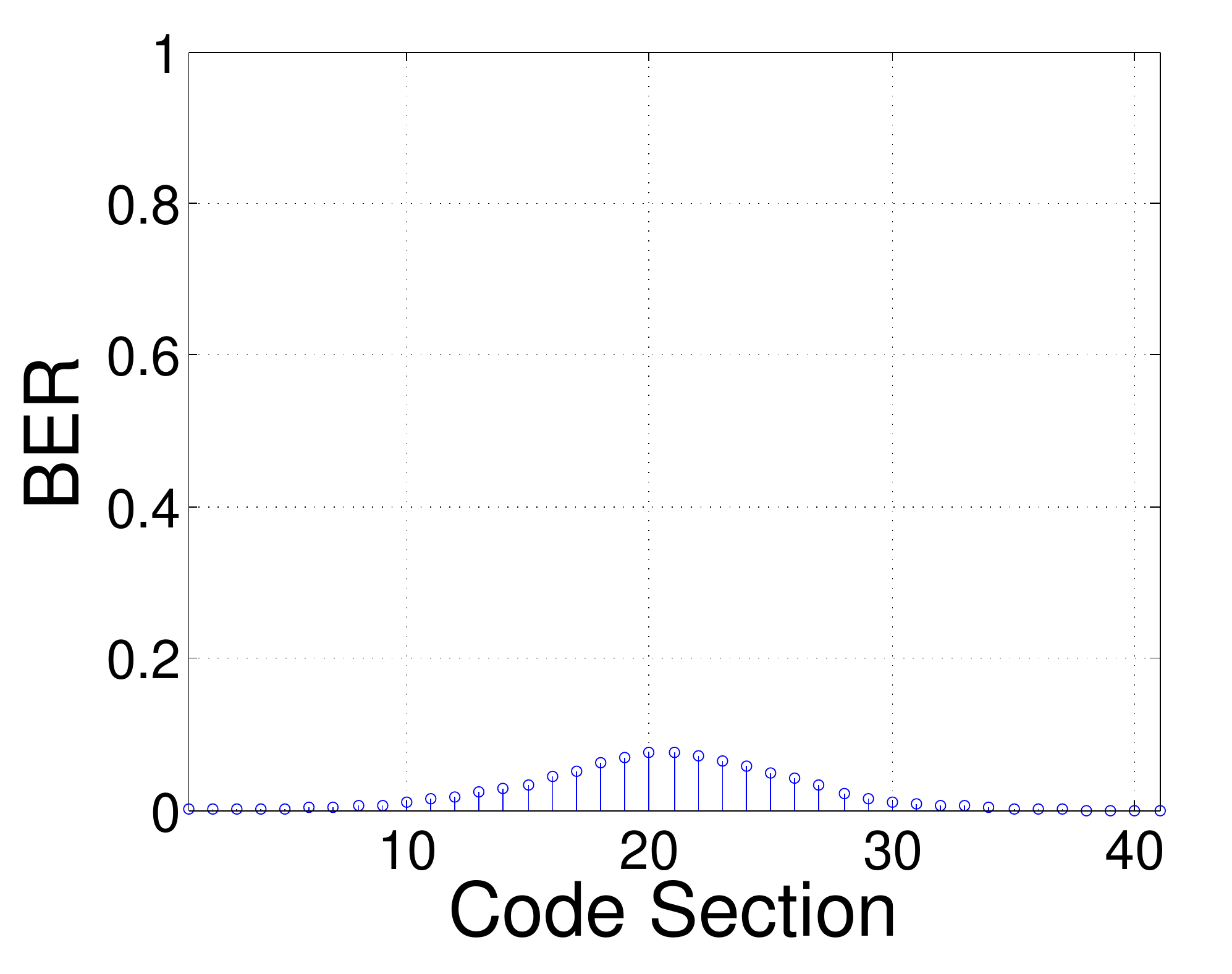}\\
(c) & (d)
\end{tabular}
\caption{Bit Error Probability per code section during the decoding process for $\ell=5$ (a), $\ell=30$ (b), $\ell=70$ (c) and $\ell=110$ (d) for a convolutional code with $L=20$, $M=1024$ and $\epsilon=0.44$. }\LABFIG{BERSEC}
\end{figure}

On the other hand, for $\pe\geq\epsilon^{\text{BP}}(l, r)$, we expect that
the number of required iterations scales linearly in $L$: the ``decoding
wave'' starts at the boundaries and moves at a constant speed towards
the middle \cite{Lentmaier10}. This can be seen in Fig. \FIG{BERSEC},
where we plot the bit error rate (BER) measured in each section of a
$(l=3,r=6,L=20,M=1024)$ ensemble after $\ell=5$ (a), $\ell=30$ (b),
$\ell=70$ (c) and $\ell=110$ (d) iterations for a channel parameter of
$\pe=0.44$. 
In Fig. \FIG{IterConv} we plot $\varphi(\ell,\epsilon,L,M)$ for
$L=5,10,20$, $M=256,512$ and $\epsilon=0.44$. We have averaged over $50$
code samples. First observe that the distribution moves to the right with
$L$ and we can see that the mean of the distributions scales linearly
in $L$ -- so the larger $L$ the more iterations we need. Further, as $M$
increases, the distribution concentrates around its mean. This means that
for large $M$ most instances decode with a number of iterations which is
close to the expected value. However, the distributions are heavy-tailed.
I.e., over a large interval the curves are approximately straight lines,
which indicates that over this range they follow a power law, i.e.,
they have the form $\ell^{\alpha}\beta$ for some non-negative constant
$\alpha$ and $\beta$. Operationally this means that, with non-negligible
probability, an instance takes  many more iterations to decode as it is
typical.  The last two conclusions are quite similar to what can be
observed for standard LDPC ensembles, see \cite{Urbanke08-2}.
One strategy to deal with the linear increase of the decoding complexity
for very large chain lengths is the application of a windowed decoder
\cite{Papaleo10}.

\begin{figure}[h]
\centering
\begin{tabular}{c}
\includegraphics[width=7.5cm,height=6.25cm]{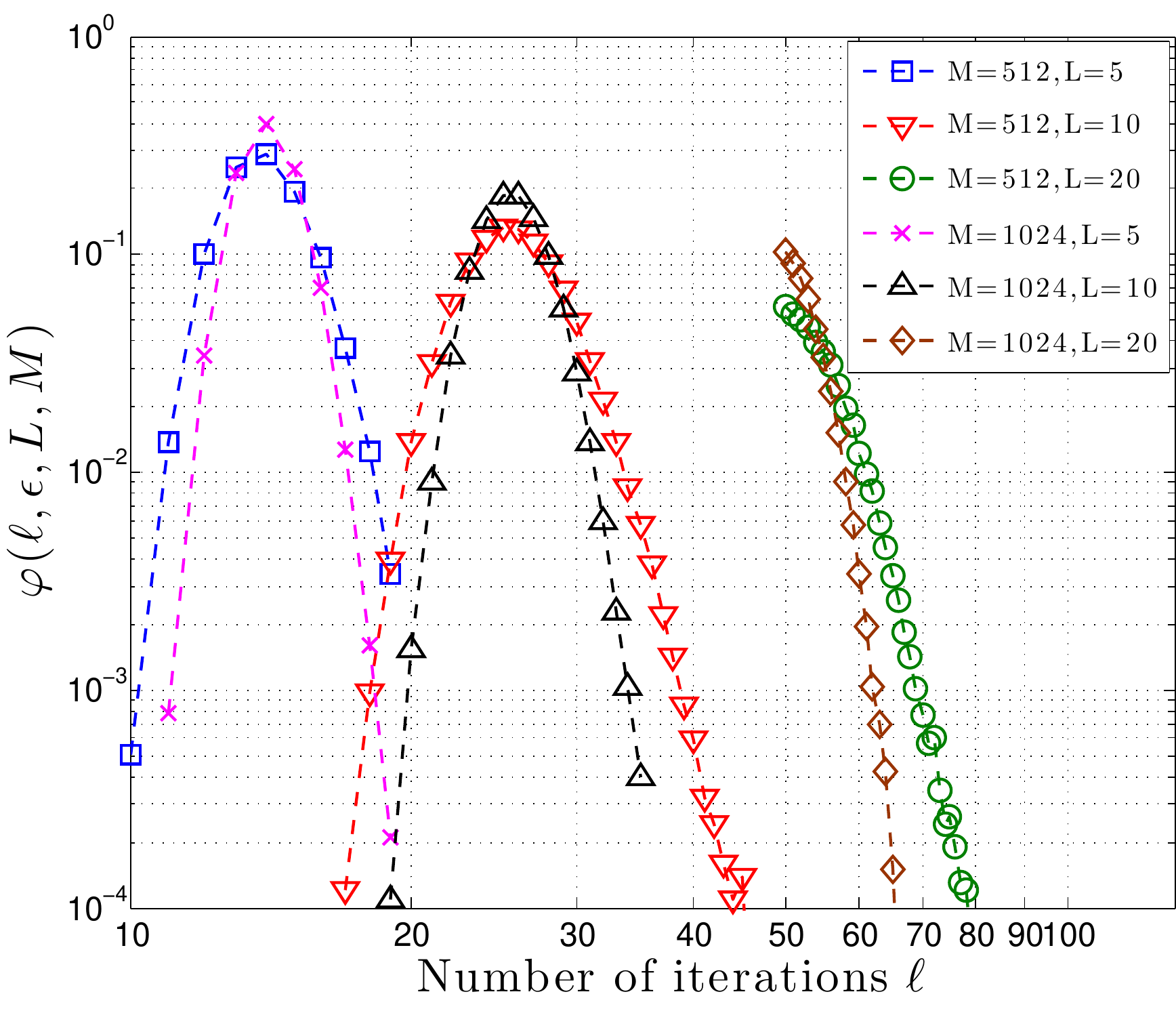} 
%\\(b)
\end{tabular}\caption{Distribution for the BP required number of iterations, for a regular (3,6) code for $L\in\{5,10,20\}$ and $M\in\{512,1024\}$ and $\epsilon=0.465$.}\LABFIG{IterConv}
\end{figure}

\section{Scaling Behavior}\LABSEC{Thbeha}
Let us now investigate for what scalings $L=f(M)$ the threshold
saturation effect appears.  We have run a large set of simulations for
various scaling functions $f(M)$ and a regular $(l,r)=(3,6)$ code has
been used to construct the ensembles. The BP and MAP thresholds for this
ensemble are, respectively, $\epsilon^{\text{BP}}(l,r)\approx0.4294$ and
$\epsilon^{\text{MAP}}(l,r)\approx0.4815$. The BP decoder is run until
all messages have converged (which always happens for the BEC).  We choose a
$(l,r)=(3,6)$ code to better illustrate the effect of the error floor in
the scaling since regular codes with larger degrees, e.g., a $(5,10)$ regular
ensemble, have much lower error floors.  Our current aim however is not to
construct optimal codes but to illustrate some typical effects.
For each $(L,M)$ pair we only consider one single sample, randomly taken
from the ensemble, as described in Section\SEC{Bpar}.  For each $\epsilon$
value and fixed code, we consider $10^{5}$ transmitted codewords.
%Standard concentration theorems ensemble average \cite{Urbanke01-2}
%assert that the performance that we get from a particular sample
%$\mathcal{C}\in(l,r,L,M)$ can be representative of the average ensemble,
%if the code length ($M$ for us) is large enough.

\subsection{Fixed $L$, Increasing $M$}
Consider first the case of constant $f(M)$. Since this is the regime
used for the DE analysis, we know that the limiting performance
$M\rightarrow\infty$ is given by \EQ{l1}. We can get a negligible
error probability as long as we are operating below the threshold
$\epsilon^{\text{MAP}}(l,r)$.  In Fig. \FIG{FixL}, we represent the bit
error probability when $L$ is fixed to $100$. As expected, the curves
become steeper as we increase $M$. Note that the curves show an error
floor.  As we discuss in more detail in Section\SEC{errorfloor}, this
error floor is due to the fact that the ratio $L/M^{l-2}$ is relatively
large for most of these cases.

%In a practical scenario, large $L$ means an increase in the decoding complexity.  Therefore, we often set $L$ as small as possible, constrained only by the fact that a small $L$
%implies a large loss in rate
\begin{figure}
\centering
\begin{tabular}{c}
\includegraphics[width=7.75cm,height=6.5cm]{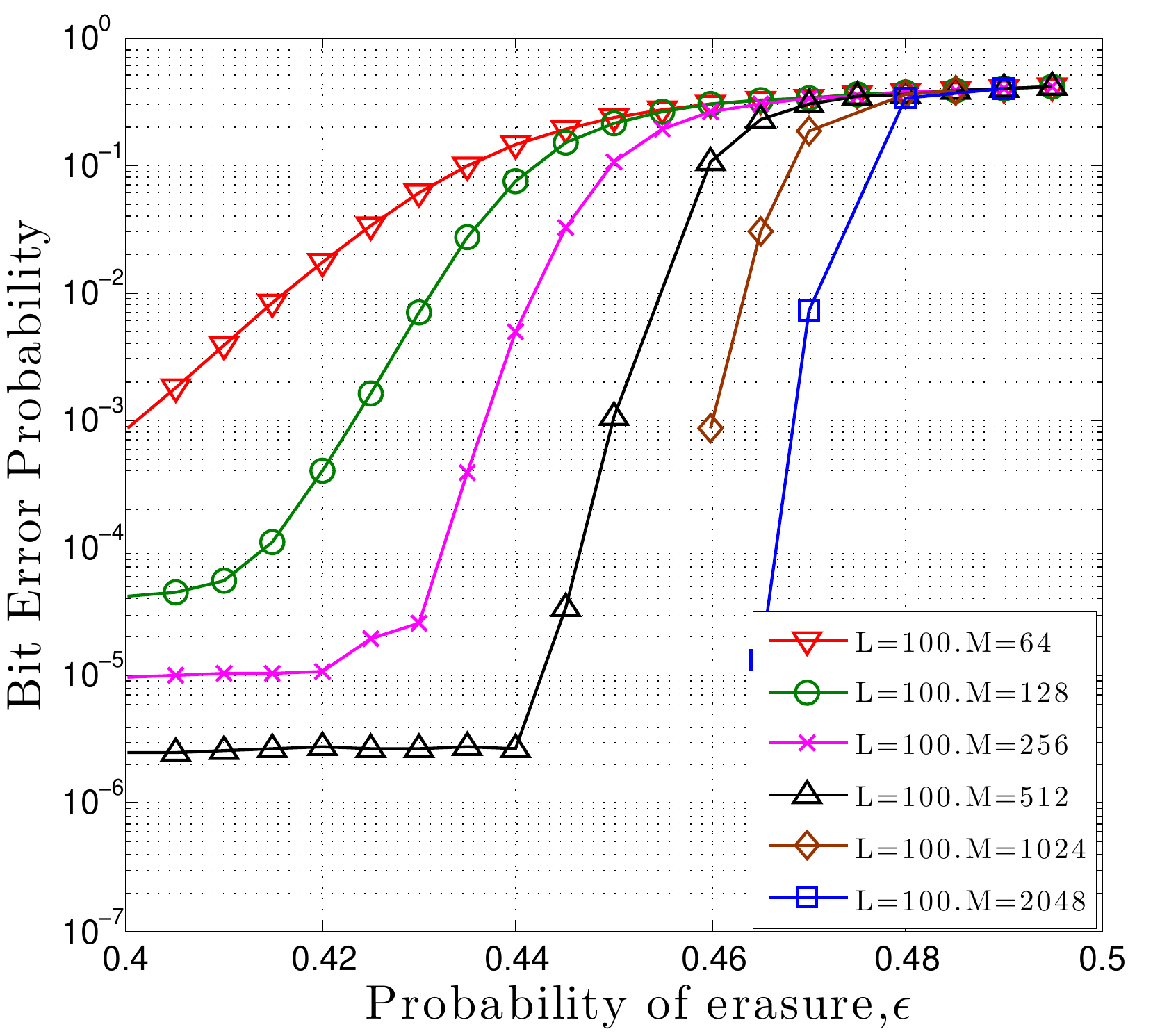} \\
\end{tabular}\caption{Bit Error probability $P_{b}^{\ell}(\epsilon,l,r,L)$ for a $(l=3,r=6,L,M)$ code. $L$ is fixed to 100.}\LABFIG{FixL}
\end{figure}

\subsection{Fixed $M$, Increasing $L$}\LABSEC{FixedM}
Let us now look at the other extreme. i.e., we fix $M$ to some
constant $M_{0}>0$ and let $L$ grow. Clearly, in this regime we do
not expect to see the same threshold saturation phenomenon.
If we consider $P_{b}^{\ell=\infty}(\epsilon,l,r,L,M_{0})$ and if we increase $L$ then we expect
this error probability to be monotonically increasing in $L$ since
the longer the chain the higher the chance that the ``decoding wave'' 
%which starts at the boundary 
gets stuck before reaching the middle.
In Fig. \FIG{FixedM256}, we plot the error probability for the case $M_{0}=512$.
As expected, the error probability is indeed monotonically
increasing in $L$ and it seems to converge to a limiting
curve.  The determination of this limiting curve is an interesting
open problem. 

\begin{figure}[h]
\centering
\includegraphics[width=7.75cm,height=6.5cm]{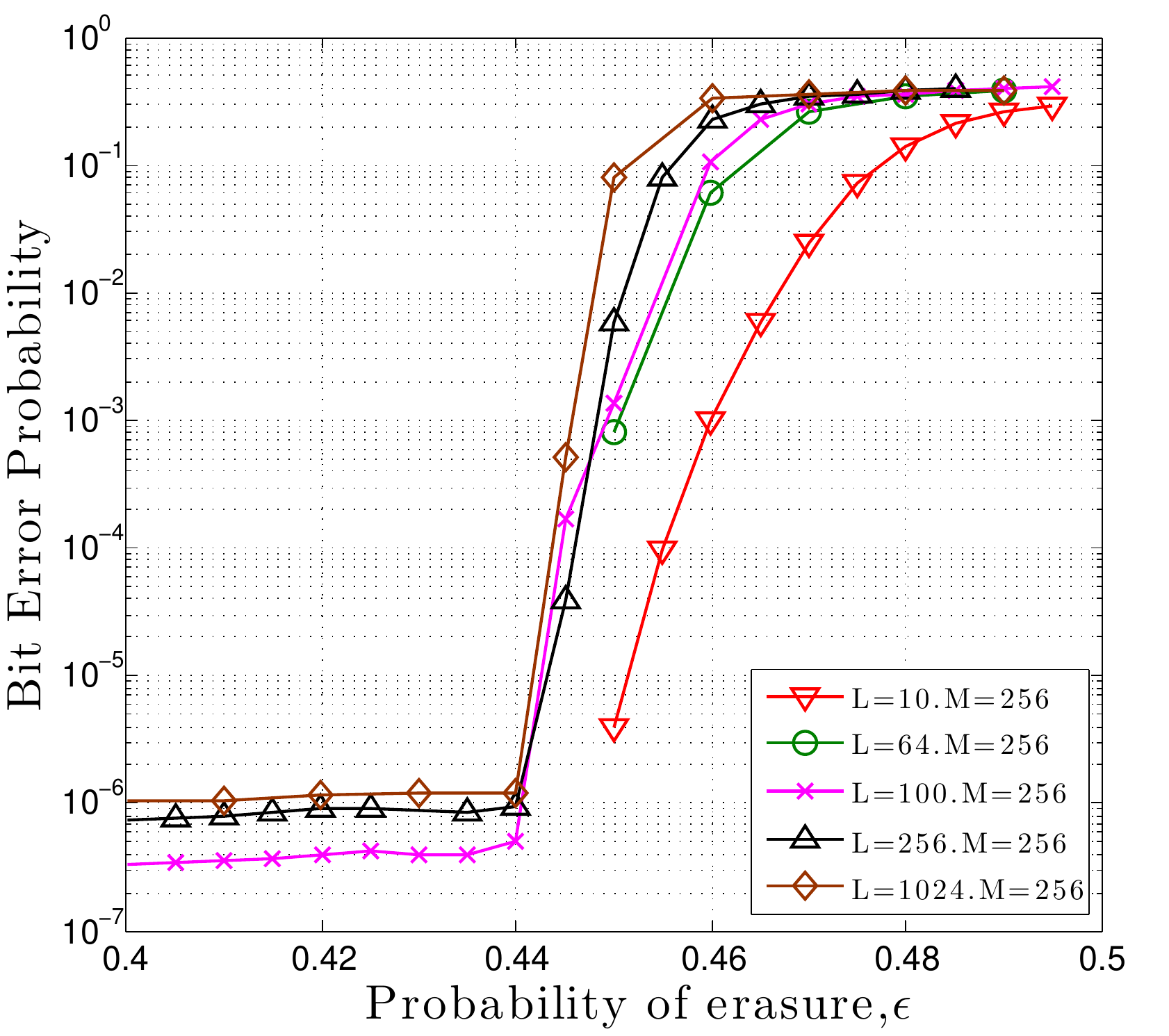} \caption{Bit Error
probability $P_{b}^{\ell}(\epsilon,l,r,L)$ for a $(l=3,r=6,L,M)$
code. $M$ is fixed to 256.}\LABFIG{FixedM256} \end{figure}

\subsection{$L$ as a general function of $M$}
Now where we have investigated the two limiting cases it is of interest
to scale {\em both} $M$ and $L$ together. At what scaling does the
behavior change? In Fig. \FIG{LigualM} and Fig. \FIG{L2}, we test the
scaling functions $L=M/2$ and $L=(M/2)^2$. In Fig. \FIG{L2}, we have
included, in dashed lines, the asymptotic ensemble error floor derived
in Section\SEC{errorfloor}. For both scaling functions the performance
improves with $M$, although in Fig.~\FIG{L2} the speed of improvement is
slower. Indeed, it seems that in both cases the asymptotic threshold
still is $\epsilon^{\text{MAP}}(l,r)$. This illustrates that
the threshold saturation phenomenon is quite robust and general.
Due to the large $L/M^{l-2}$ values, we observe large error floor
levels in both cases.  Finally, in Fig.~\FIG{Expv1} we plot the
performance of an extreme scenario, where $L$ scales exponentially
with $M$.  The performance now worsens with $L$, similarly to the case
considered in Section\SEC{FixedM}. A back of the envelope calculation,
proposed to us by Andrea Montanari, suggests that an exponential scaling
relationship is exactly the boundary -- for a subexponential growth of
$L$ as a function of $M$ we expect the threshold phenomenon to happen
whereas for super-exponential growths we expect it not to occur.

%
% i.e., if $L$ grows subexponentially
%in $M$ one would expect the threshold saturation phenomenon to occur
%but not when $L$ grows faster than exponential.

\begin{figure}
\centering
\includegraphics[width=8cm,height=6.5cm]{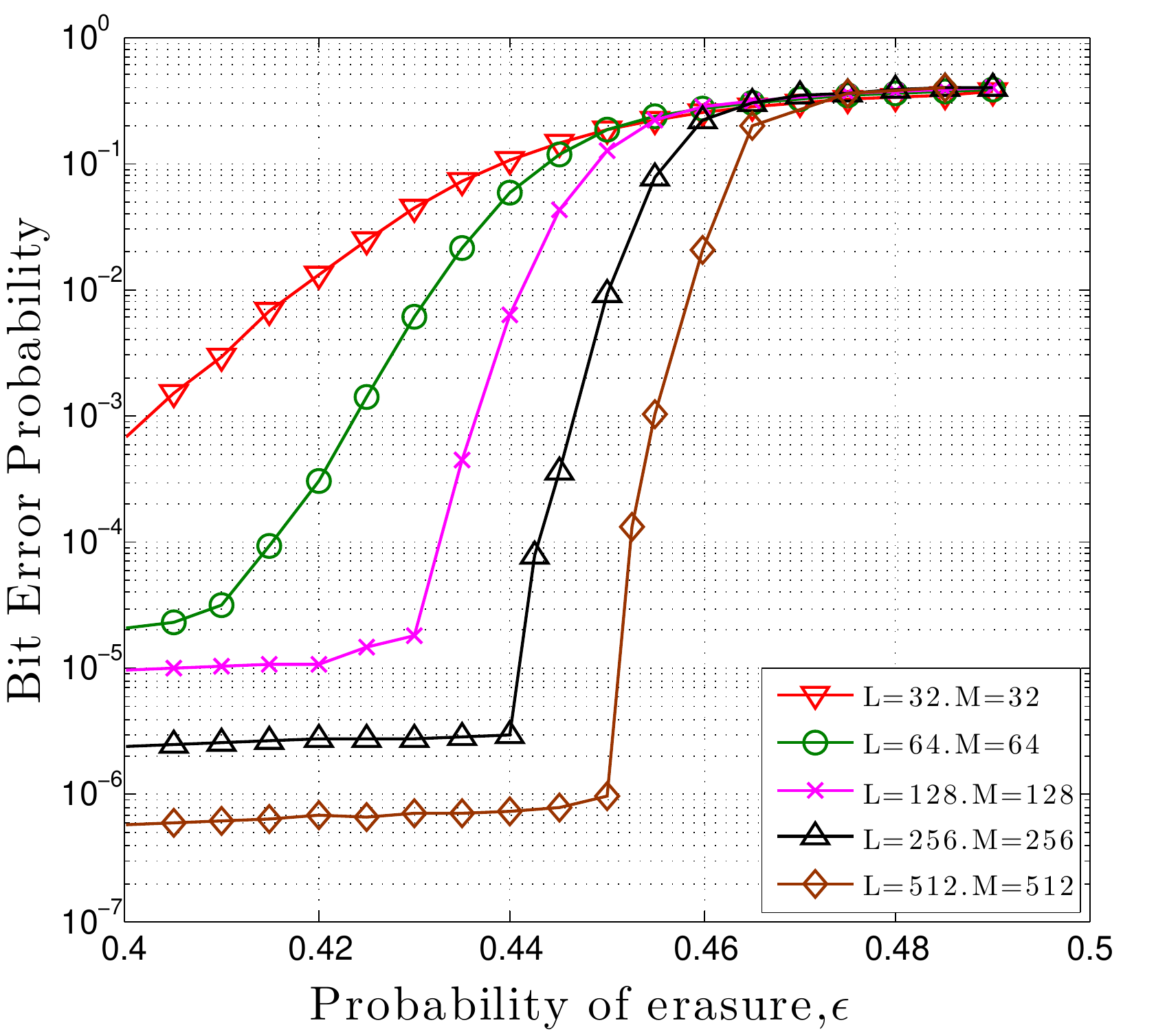} \caption{Bit Error probability $P_{b}^{\ell}(\epsilon,l,r,L)$ for a $(l=3,r=6,L,M)$ code. $L$ is equal to $M/2$.}\LABFIG{LigualM}
\end{figure}

\begin{figure}
\centering
\includegraphics[width=7.75cm,height=6.5cm]{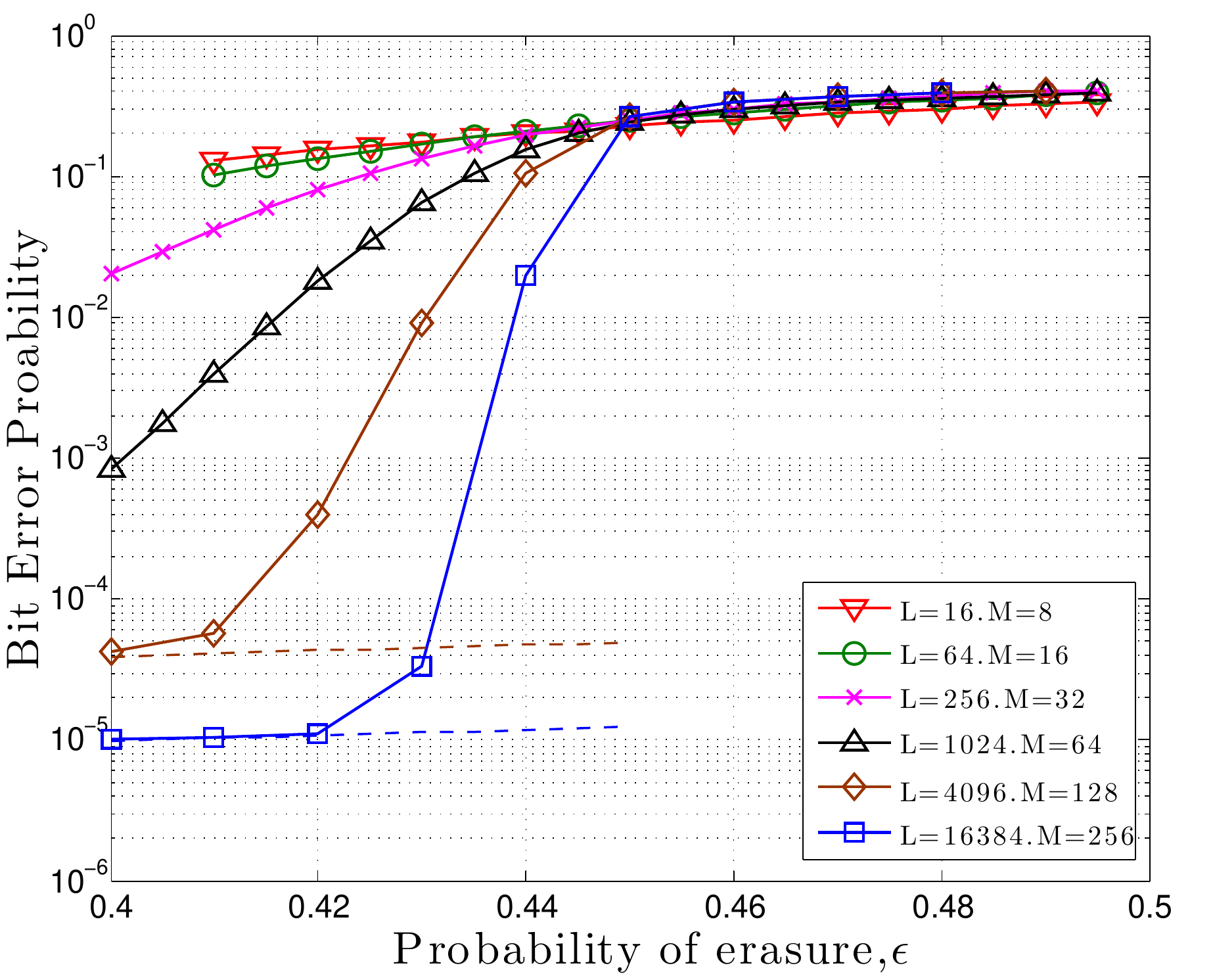} \caption{Bit Error probability $P_{b}^{\ell}(\epsilon,l,r,L)$ for a $(l=3,r=6,L,M)$ code. $L$ is equal to $(M/2)^{2}$. In dashed lines, we represent the asymptotic ensemble error floor in \EQ{ExBER} for $M=128$ and $M=256$. }\LABFIG{L2}
\end{figure}

\begin{figure}
\centering
\includegraphics[width=7.75cm,height=6.5cm]{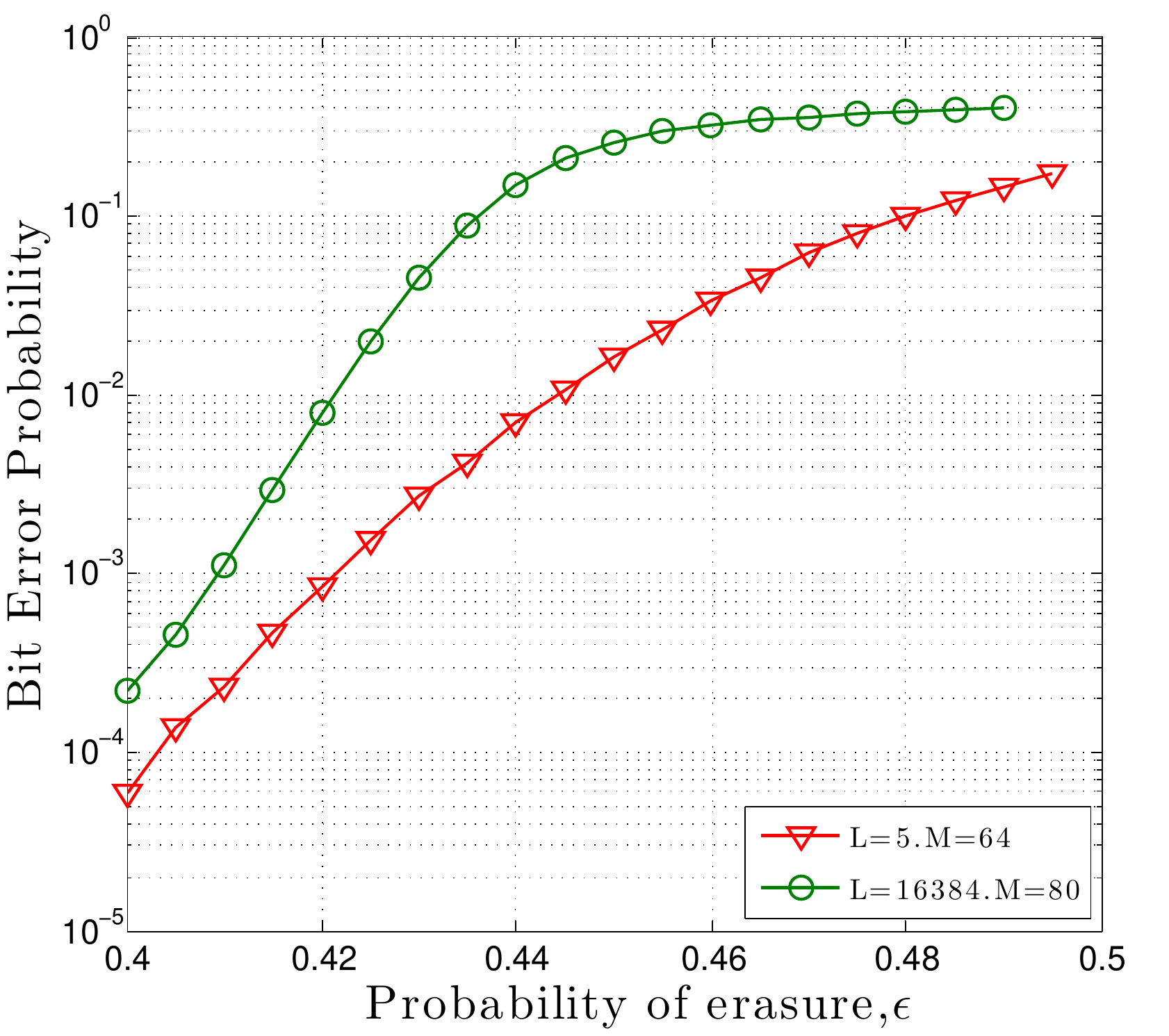} \caption{Bit Error probability $P_{b}^{\ell}(\epsilon,l,r,L)$ for a $(l=3,r=6,L,M)$ code. $L$ scales exponentially with $M$.}\LABFIG{Expv1}
\end{figure}
Let us summarize. The threshold saturation phenomenon happens empirically
over a very wide range of scalings of $L$ with respect to $M$ and far
beyond what theory currently can predict. This is of comfort to the code
designer in the field and a challenge for any theoretician.

\section{Error Floor}\LABSEC{errorfloor}
In many of the former simulations, we have seen the occurrence of error
floors.  Let us now quickly discuss how this error floor can be analyzed.
To simplify matters, we only consider codewords/stopping sets of weight two
since codewords/stopping sets of higher weight vanish (in $M$) at a much
higher rate.

\lemma[Convergence to Poisson Distribution]{\label{lem:weight} Consider an LDPCC ensemble $(l,r=kl,L,M)$. Let  $\mathcal{C}$ be a code sample and $\ntwo$ be the number codewords with Hamming weight two in $\mathcal{C}$. 
Assume that the code is chosen randomly with a uniform probability from the ensemble. 
Then the distribution of $\ntwo$ converges (in $M$) to 
\begin{align}\LABEQ{Poisson}
&\ntwo\sim \textrm{Pois}\left(\lambda\right),\quad \lambda=k^{l-2}\binom{k}{2}\frac{(2L+1)}{M^{l-2}}\quad k\geq2.
\end{align}\LABLEM{Le1}}
\begin{proof}
%Let us just sketch the proof. 
Note that a codeword of weight two is only
formed by two variables in the same section, see Fig. \FIG{Fig2}, that share the same set of $l$ check nodes. Since there
are $(M/k)$ check nodes per section, this happens with probability
$p=(M/k)^{-l}$. In each section, we count $\binom{k}{2}(M/k)^{2}$ pairs of variables that can form a weight two codeword. Therefore, in a graph with $(2L+1)$ sections,
the expected number of such codewords converges to $\lambda=(2L+1)\binom{k}{2}(M/k)^{2}p$. That the distribution converges
to a Poisson distribution follows by standard arguments as in the case of uncoupled
LDPC ensembles, see \cite{Urbanke08-2}.
\end{proof}

\corollary[Fraction of Codes with No Small Codewords]{The fraction of codes in the $(l,r=kl,L,M)$ ensemble
with no codewords of weight $2$ converges to $\exp(-\lambda)$.} 
\begin{proof}
The expected fraction of such codes
is given by $\P(\ntwo=0)$, which is $\exp(-\lambda)$
by Lemma\LEM{Le1}.
\end{proof}

The accuracy of Lemma~\ref{lem:weight} is illustrated in Fig.~\FIG{Pois},
where we compare the Poisson distribution in \EQ{Poisson} with some
experimental data, obtained by analyzing $10^{4}$ code samples. We 
consider  an $(l=3,r=6,L=100,M=128)$ ensemble and we plot the
experimental normalized histogram $(\circ)$ along with the Poisson distribution
in \EQ{Poisson} $(\ast)$. We can see that both plots fit almost perfectly.

\begin{figure}[h]
\centering
\includegraphics[width=7cm,height=6cm]{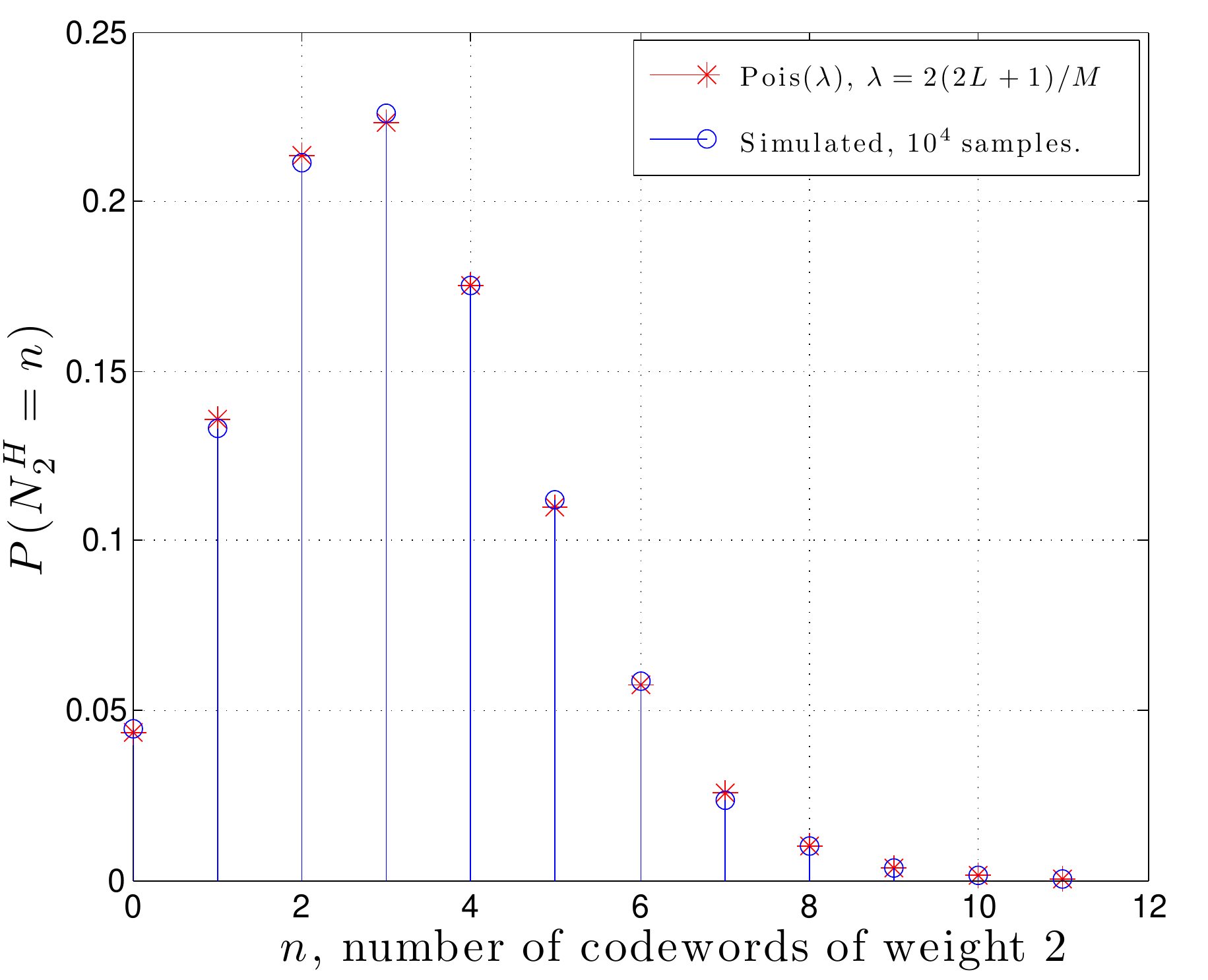} \caption{Poisson distribution approximation for $\ntwo$ $(\ast)$ and experimental estimation of the pdf $(\circ)$ for an ensemble $(l=3,r=6,L=100,M=128)$.}\LABFIG{Pois}
\end{figure}

\corollary{The expected error floor of an $(l,r,L,M)$ ensemble is given by
\begin{equation}\LABEQ{ExEF}
P_{b}^{\ell}(\epsilon,l,r,L,M)=2\binom{k}{2}k^{l-2}\frac{\epsilon^{2}}{M^{l-1}},\;\;\epsilon\ll\epsilon^{\text{BP}}(l,r,L).
\end{equation}
}

\begin{proof}
In the error floor region, we compute the BER as follows:
\begin{align}\LABEQ{ExBER}
P_{b}^{\ell}(\epsilon,l,r,L,M)&=
\E_{\mathcal{C}\in(l,r,L,M)}\left[P_{b}^{\ell}(\epsilon,\mathcal{C})\right]\nonumber
\stackrel{(a)}{=}\E_{\mathcal{C}}\left[\frac{2\ntwo\epsilon^{2}}{M(2L+1)}\right]\nonumber\\&=\frac{2\lambda\epsilon^{2}}{M(2L+1)}=2\binom{k}{2}k^{l-2}\frac{\epsilon^{2}}{M^{l-1}},
\end{align}
where, in step $(a)$, we have assume that the error floor is due to
codewords of weight two. A given sample $\mathcal{C}$ has $\ntwo$
of such codewords and in
average, $\epsilon^{2}\ntwo$ of them are erased.  
\end{proof}

From the above observations we can deduce the following.  For a particular
scaling of $L=f(M)$, if $\lambda$ stays bounded from above by a small
constant or even tends to $0$, then it is easy to expurgate the ensemble
and hence to avoid error floors.  This always happens if $L$ grows slower
than $M^{l-2}$, a condition which is easy to achieve in practice. In order
to illustrate the accuracy of analytical error floor predictions we have on
purpose considered ensembles that are hard
to expurgate, i.e., for these ensembles most code samples have error
floor, which is predicted by \EQ{ExEF}. For instance, in Fig. \FIG{L2},
we have plotted in dashed lines the error floor in \EQ{ExEF} for the
cases $(M=128,L=4096)$ and $(M=256,L=16384)$, where we can observe the
accuracy of the estimate.

%{\small
%\section*{Acknowledgement}
%This work was supported by grant No. 200021-125347 of the Swiss
%National Foundation and by Spanish government MEC TEC2009-14504-C02-\{01,02\} and Consolider-Ingenio 2010 CSD2008-00010).
%}

%Let us finally compute the variance of the error floor that a particular
%code samples presents with respect the average in \EQ{ExBER}:
%\begin{align}\LABEQ{Var}
%\sigma^{2}_{\mathcal{C}}(l,r,L,M)=&\E\left[\left(\frac{2\epsilon^{2}\ntwo}{2M(2L+1)}-\frac{2\lambda\epsilon^{2}}{2M(2L+1)}\right)^{2}\right]\\
%&=\frac{\epsilon^{4}}{M^{2}(2L+1)^{2}}\sigma^{2}_{\ntwo}=\frac{\epsilon^{4}}{M^{l}(2L+1)(\frac{2}{k})^{l}}\nonumber,
%\end{align}
%where $\sigma^{2}_{\ntwo}$ is the variance of $\ntwo$, equal
%to $\lambda$ since it follows a Poisson distribution. Note that
%$\sigma_{\mathcal{C}}(l,r,L,M)$ quickly tends to zero as $M$ and $L$ grow,
%which explain the similar error floor levels in Fig.  \FIG{LigualM} and
%Fig. \FIG{L2}. Indeed, they perfectly match with the average in \EQ{ExEF}.

\bibliography{LDPC,Gmodels,LDPC_Conv}
\bibliographystyle{IEEEtran}

\appendices
\end{document}